\documentclass[twocolumn, pra]{revtex4-1}

\usepackage{natbib}
\usepackage{graphicx}
\usepackage{dcolumn} 
\usepackage{bm}
\usepackage{hyperref}
\usepackage{mathtools}
\usepackage{amssymb}
\usepackage{enumitem}
\usepackage[mathlines]{lineno}
\usepackage[titletoc, title]{appendix}

\begin{document}

\title{Pathway toward the formation of supermixed states in ultracold boson mixtures loaded in ring lattices }

\author{Andrea Richaud}
\email{andrea.richaud@polito.it}
\author{Vittorio Penna}
\affiliation{
Dipartimento di Scienza Applicata e Tecnologia and u.d.r. CNISM, Politecnico di Torino, 
Corso Duca degli Abruzzi 24, I-10129 Torino, Italy}
\date{\today}

\begin{abstract}
We investigate the mechanism of formation of supermixed soliton-like states in bosonic binary mixtures loaded in ring lattices. We evidence the presence of a common pathway which, irrespective of the number of lattice sites and upon variation of the interspecies attraction, leads the system from a mixed and delocalized phase to a supermixed and localized one, passing through an intermediate phase where the supermixed soliton progressively emerges. The degrees of mixing, localization and quantum correlation of the two condensed species, quantified by means of suitable indicators commonly used in Statistical Thermodynamics and Quantum Information Theory, allow one to reconstruct a bi-dimensional mixing-supermixing phase diagram featuring two characteristic critical lines. Our analysis is developed both within a semiclassical approach capable of capturing the essential features of the two-step mixing-demixing transition and with a fully-quantum approach.

\end{abstract}

\maketitle

\section{Introduction}
\label{sec:Introduction}
In the last decade, a considerable attention has been paid to the mixing-demixing transitions occurring in bosonic binary mixtures confined in optical lattices. Such systems, realized by means of both homonuclear \cite{Soltan_2} and heteronuclear \cite{Catani_deg} components, show how the interplay among the \textit{intra-}species and the \textit{inter-} species repulsion, the tunneling effect and the fragmentation induced by the periodic potential strongly affects the mixing properties and gives rise to an extremely rich phenomenology. This includes spatial phase separation in large-size lattices \cite{sep1,sep3}, mixing properties of dipolar bosons \cite{sep2}, quantum emulsions \cite{Buonsante_PRL,qe1}, the structure of quasiparticle spectrum across the demixing transition \cite{Angom}, and the influence on phase separation of thermal effects \cite{Roy}, interspecies entanglement \cite{ent}, and asymmetric boson species \cite{Belemuk}. Further aspects concerning the interlink between demixing and the dynamics of mixtures have been explored in \cite{Kasa,Gallemi_1,tic,Noi_NJP}.

Recently, spatial phase separation has been investigated for repulsive interspecies interactions in small-size lattices \cite{Bruno,NoiEntropy,NoiSREP,Noi_Zenesini}. This analysis has disclosed an unexpectedly-complex demixing mechanism in which the regimes with fully-separated and the fully-mixed components are connected by an intermediate phase still exhibiting partial mixing.      
Overall, the resulting phases feature specific miscibility properties which can be quantified by means of the entropy of mixing, an indicator originally introduced in the context of macromolecular simulations \cite{Camesasca}. The demixing of two quantum fluids, and their ensuing localization in different spatial domains, has been shown to be strictly linked with the presence of criticalities in several quantum indicators including, but not limited to, ground-state energy, energy levels' structure and entanglement between the species \cite{NoiSREP,PennaLinguaJPB}.  

In this work, we aim at exploring the characteristic regimes of the mixture when the interaction between the condensed species is \textit{attractive}. The competition between the interspecies \textit{attraction} and the intraspecies \textit{repulsions} results in a rather rich variety of phenomena which culminates in the formation of a \textit{supermixed} soliton, i.e. a configuration where both condensed species localize in a unique site. 

The scope of our analysis is rather broad, both because we take into account the possible presence of asymmetries between the condensed species and because the analysis itself is developed for a generic $L$-site trapping potential with ring geometry. A semiclassical scheme based on the approximation of inherently discrete quantum numbers with continuous variables (hence the name ``continuous variable picture" (CVP)) allows one to reduce the original quantum problem to a classical one \cite{Spekkens,Javanainen,Ciobanu,Zin,Penna_Burioni}. The latter, in turn, displays, in a rather transparent way, the occurrence of critical phenomena such as the formation of soliton-like configurations and the onset of mixing-demixing or mixing-supermixing transitions.

Interestingly, our analysis not only highlights the fact that the formation of a supermixed soliton constitutes a two-step process, made possible by the non-linearity of the interspecies-attraction term, but also that this two-step process occurs in a generic $L$-site potential, no matter the specific value of $L$. In other words, depending on the strength of the interspecies attraction, but irrespective of the value of $L$, the system's ground state exhibits three qualitatively different spatial structures: \textit{i)} the one featuring uniform boson distribution among all the wells, \textit{ii)} the one already including the seed of the supermixed soliton but featuring an incomplete localization and \textit{iii)} the one where the supermixed soliton is fully emerged and developed. 

The phase diagram derived within this semiclassical approach is then validated by means of several genuinely quantum indicators, which indeed confirm the presence of three qualitatively different classes of ground states and the occurrence of a two-step process leading to the formation of supermixed solitons.

The outline of this manuscript is the following: in Sec. \ref{sec:Model} we present the quantum model for a bosonic binary mixture confined in a $L$-site potential and its semiclassical approximation. In Sec. \ref{sec:Phase_diagram} we present the system's phase diagram when the boson populations tends to infinity. Note that this circumstance can be interpreted as a well-defined thermodynamic limit, in the sense of the statistical-mechanical approach developed in \cite{Dynamical_Bifurcation,Oelkers}. We also introduce two indicators which can be conveniently used to quantify the degree of mixing and of localization of the two quantum fluids. Sec. \ref{sec:Delocalizing_effect} is devoted to the analysis of the system's properties for finite hopping amplitudes. In Sec. \ref{sec:Quantum_indicators} we present a number of quantum indicators whose critical character corroborates the discussion developed in the previous sections. Eventually, Sec. \ref{sec:Conclusions} is devoted to concluding remarks.

\section{The model}
\label{sec:Model}
\subsection{The quantum model}
In this article, we focus on the supermixing effect and on the soliton-formation mechanism in a two-component bosonic mixture loaded in $L$-site potentials. The genuinely quantum features of such system can be effectively captured by the second-quantized Hamiltonian   
$$
H= - T_a \sum_{j=1}^{L} \left(a_{j+1}^\dagger a_j +a_j^\dagger a_{j+1} \right) + \frac{U_a}{2} \sum_{j=1}^{L} n_j(n_j-1)
$$
$$
  - T_b \sum_{j=1}^{L} \left(b_{j+1}^\dagger b_j +b_j^\dagger b_{j+1} \right)+\frac{U_b}{2} \sum_{j=1}^{L} m_j(m_j-1)
$$
\begin{equation}
\label{eq:BH}
		  +W \sum_{j=1}^{L} n_j\, m_j, 
\end{equation}
an extended version of the well-known Bose-Hubbard model whose last term accounts for the \textit{attractive} interaction between the species. Operator $a_i$ $(b_i)$ destroys a species a (species b) boson in the $i-$th site. Notice that $i\in\{1,\,\dots,\,L\}$ and that, for $L>2$, the trapping potential is assumed to feature a ring geometry, a circumstance which results in the periodic boundary conditions $i=L+1\equiv 1$. As a consequence of the bosonic character of the trapped particles, the following commutation relations hold: $[a_i,b_\ell^\dagger]=0$, $[a_i,a_\ell^\dagger]=[b_i,b_\ell^\dagger]=\delta_{i,\ell}$. The definition of number operators, $n_i=a_i^\dagger a_i$ and $m_i=b_i^\dagger b_i$, allows one to evidence two independent conserved quantities, namely $N_a=\sum_i^L n_i$ and $N_b=\sum_i^L m_i$. Concerning model parameters, $T_a$ and $T_b$ represent the tunnelling energy of the two species, $U_a>0$ and $U_b>0$ their \textit{intra-}species \textit{repulsive} interactions, and $W<0$ the \textit{inter-}species \textit{attractive} coupling.      

\subsection{A Continuous Variable Picture for the detection of different quantum phases}
\label{subsec:CVP}
An effective way to determine the ground state structure of multimode BH Hamiltonians consists in approximating the inherently discrete single-site occupation numbers $n_j$ and $m_j$ with continuous variables $x_j=n_j/N_a$ and $y_j=m_j/N_b$ \cite{Spekkens,Ciobanu,PennaLinguaPRE,Dynamical_Bifurcation,NoiSREP,Noi_Zenesini}. Provided that the overall boson populations, $N_a=\sum_j n_j$ and $N_b=\sum_j m_j$, are large enough, it is in fact possible to establish a one-to-one correspondence between a certain Fock state $\left| n_1,\,\dots,\, n_L,\,m_1,\,\dots,\, m_L \right\rangle =: |\vec{n},\,\vec{m}\rangle$ and state $\left|x_1,\,\dots,\, x_L,\,y_1,\,\dots,\, y_L \right\rangle =: |\vec{x},\,\vec{y}\rangle $, i.e. to turn \textit{integer} quantum numbers $n_j$ and $m_j$ into \textit{real} variables $x_j$ and $y_j$, both $\in[0,1]$. With this in mind, creation and annihilation processes $n_j\to n_j\pm 1$ $(m_j\to m_j\pm 1)$ can be associated to small changes of the corresponding continuous variable, i.e. $x_j\to x_j\pm \epsilon_a$ $(y_j\to y_j\pm \epsilon_b)$ where $\epsilon_a=1/N_a \ll 1$ $(\epsilon_b=1/N_b \ll 1)$. In the following, therefore, we will focus on those regimes where the total number of atoms is large enough to justify the use of the CVP, but low enough not to break the tight-binding approximation required to obtain the single-band Bose-Hubbard Hamiltonian (\ref{eq:BH}). The application of this approximation scheme to a second-quantized Hamiltonian of the type (\ref{eq:BH}) allows one to reformulate it in terms of generalized coordinates $x_j$ and $y_j$ and of their conjugate momenta. As a consequence, within such scheme, the (quadratic approximation of the) eigenvalue problem $H|E\rangle=E|E\rangle$ reads
\begin{equation}
\label{eq:Eigenvalue_problem}
 (-\mathcal{D}+\mathcal{V})\psi_E(\vec{x},\vec{y}) = E\,\psi_E(\vec{x},\vec{y})
\end{equation}
where
$$
 \mathcal{D}= -\frac{T_a}{N_a} \sum_{j=1}^L \left[\left(\partial_{x_j}-\partial_{ x_{j+1}}\right) \sqrt{x_j x_{j+1}}\left(\partial_{ x_j}-\partial_{ x_{j+1}}\right)  \right] 
$$
$$
  -\frac{T_b}{N_b} \sum_{j=1}^L \left[\left(\partial_{y_j}-\partial_{ y_{j+1}}\right) \sqrt{y_j y_{j+1}}\left(\partial_{y_j}-\partial_{ y_{j+1}}\right)  \right]
$$
is the generalized Laplacian and 
$$
   \mathcal{V}= -2N_aT_a \sum_{j=1}^L \sqrt{x_jx_{j+1}}-2N_bT_b \sum_{j=1}^L\sqrt{y_jy_{j+1}}
$$
$$
  +\frac{U_a N_a^2}{2} \sum_{j=1}^L x_j(x_j-\epsilon_a)
  +\frac{U_b N_b^2}{2} \sum_{j=1}^L y_j(y_j-\epsilon_b)
$$
\begin{equation}
\label{eq:Potenziale_lungo}
  + W N_aN_b \sum_{j=1}^L x_jy_j
\end{equation}
is the generalized potential. Provided that $\psi_E$ is well localized in the global minimum of $\mathcal{V}$ (this condition is certainly achieved if $N_a$ and $N_b$ are large enough, in that $\mathcal{D}\propto N_c^{-1} $ while $\mathcal{V}\propto N_c$, with $c=a,\,b$), potential $\mathcal{V}$ provides a remarkably effective way to investigate the ground state structure of Hamiltonian (\ref{eq:BH}) as a function of model parameters. To be more clear, the $2L$-tuples $(\vec{x},\vec{y})$ which minimize function $\mathcal{V}$ on its domain 
$$
\mathcal{R}=\left\{(\vec{x}_j,\vec{y}_j)\,:\, 0\le x_j,\,y_j\le 1, \quad \sum_{j=1}^L x_j=\sum_{j=1}^L y_j= 1 \right \}
$$
correspond to those Fock states $|\vec{n},\vec{m}\rangle$ featuring the largest weights $|c(\vec{n},\vec{m})|^2$ in the expansion of the ground state, i.e. in $|\psi_0\rangle=\sum_{\vec{n},\vec{m}}^Q c(\vec{n},\vec{m}) |\vec{n},\,\vec{m}\rangle$, where the superscript $Q$ recalls that
\begin{equation}
\label{eq:Q}
     Q=\frac{(N_a+L-1)!}{N_a!(L-1)!} \frac{(N_b+L-1)!}{N_b!(L-1)!}
\end{equation}
is the dimension of the constant-boson-number subspace contained in the Hilbert space of states associated to Hamiltonian (\ref{eq:BH}).  

The determination of the minimum points of potential $\mathcal{V}$ is of particular interest when $T_a/(U_aN_a)\to 0$ and $T_b/(U_bN_b)\to 0$. These limiting conditions, in fact, can be regarded as a sort of thermodynamic limit according to the statistical-mechanical approach discussed in \cite{Dynamical_Bifurcation,Oelkers} and, when they hold, the different phases of the quantum system (\ref{eq:BH}) emerge at their clearest \cite{NoiSREP,Noi_Zenesini}. In this limit, generalized potential (\ref{eq:Potenziale_lungo}) can be conveniently recast as
\begin{equation}
\label{eq:V}
  V\approx\frac{\mathcal{V}}{U_aN_a^2} = \frac{1}{2} \sum_{j=1}^L x_j^2 + \frac{\beta^2}{2} \sum_{j=1}^L y_j^2 + \alpha\beta \sum_{j=1}^L x_j y_j,
\end{equation}
an expressions which defines a new (rescaled) effective potential which depends only on two effective parameters
\begin{equation}
    \label{eq:alfa_beta}
    \alpha= \frac{W}{\sqrt{U_aU_b}}, \qquad \beta=\frac{N_b}{N_a} \sqrt{\frac{U_b}{U_a}}.
\end{equation}
The former constitutes the ratio between the interspecies attractive coupling and the (geometric average of) the intraspecies repulsions, while the latter corresponds to the degree of asymmetry between species a and species b condensates. Notice, in particular, that $\beta \to 1$ in the twin-species scenario, while $\beta \to 0$ when species b represents an impurity with respect to species a. In the following, we will assume $\beta \in [0,1]$ without loss of generality, as one can always swap species labels in order for $\beta$ to fall in this interval.   

Effective model parameters $\alpha$ and $\beta$ have already proved to be the most natural ones to describe the occurrence of rather complex phase-separation phenomena in ultracold binary mixtures loaded in spatially-fragmented geometries \cite{Noi_Zenesini} and, in the present case, constitute the most effective variables to capture the formation of supermixed solitons. Parameters $\alpha$ and $\beta$ span, in fact, a two-dimensional phase diagram where the various phases included therein correspond to different functional dependencies of the minimum-energy configuration $(\vec{x}_*,\vec{y}_*)$ and of the relevant energy 
\begin{equation}
\label{eq:V_*}
V_*:=V(\vec{x}_*,\vec{y}_*):=\min_{(\vec{x},\vec{y})\in \mathcal{R}}V(\vec{x},\vec{y})
\end{equation}
on $\alpha$ and $\beta$ themselves. The presence of different functional dependencies of $V_*$ on model parameters $\alpha$ and $\beta$ results in the presence of borders on the $(\alpha,\beta)$ plane where function $V_*$ is not analytic, a circumstance which strongly resembles the signature of quantum phase transitions \cite{Sachdev}.

The search for the configuration $(\vec{x}_*,\vec{y}_*)$ which minimizes function $V$ on its closed domain $\mathcal{R}$ can be carried out in a fully analytic way. Nevertheless, the complexity of such analysis increases with increasing lattice size $L$, not only because the \textit{interior} of region $\mathcal{R}$ gets bigger and bigger but also (and above all) because the \textit{boundary} of $\mathcal{R}$ gets increasingly complex and branched. Indeed, for wide regions of the $(\alpha,\beta)$ plane, it is on the boundary of $\mathcal{R}$ that $V_*$ falls, a circumstance which makes it necessary its complete exploration (see \cite{NoiSREP} for further details on the systematic analysis of the closed $(2L-2)-$polytope representing the domain $\mathcal{R}$).   

\section{The mixing-supermixing phase diagram for $T_a/(U_aN_a),\, T_b/(U_bN_b)\to 0$}
\label{sec:Phase_diagram}
\begin{figure}[h!]
\centering
\includegraphics[width=0.8\linewidth]{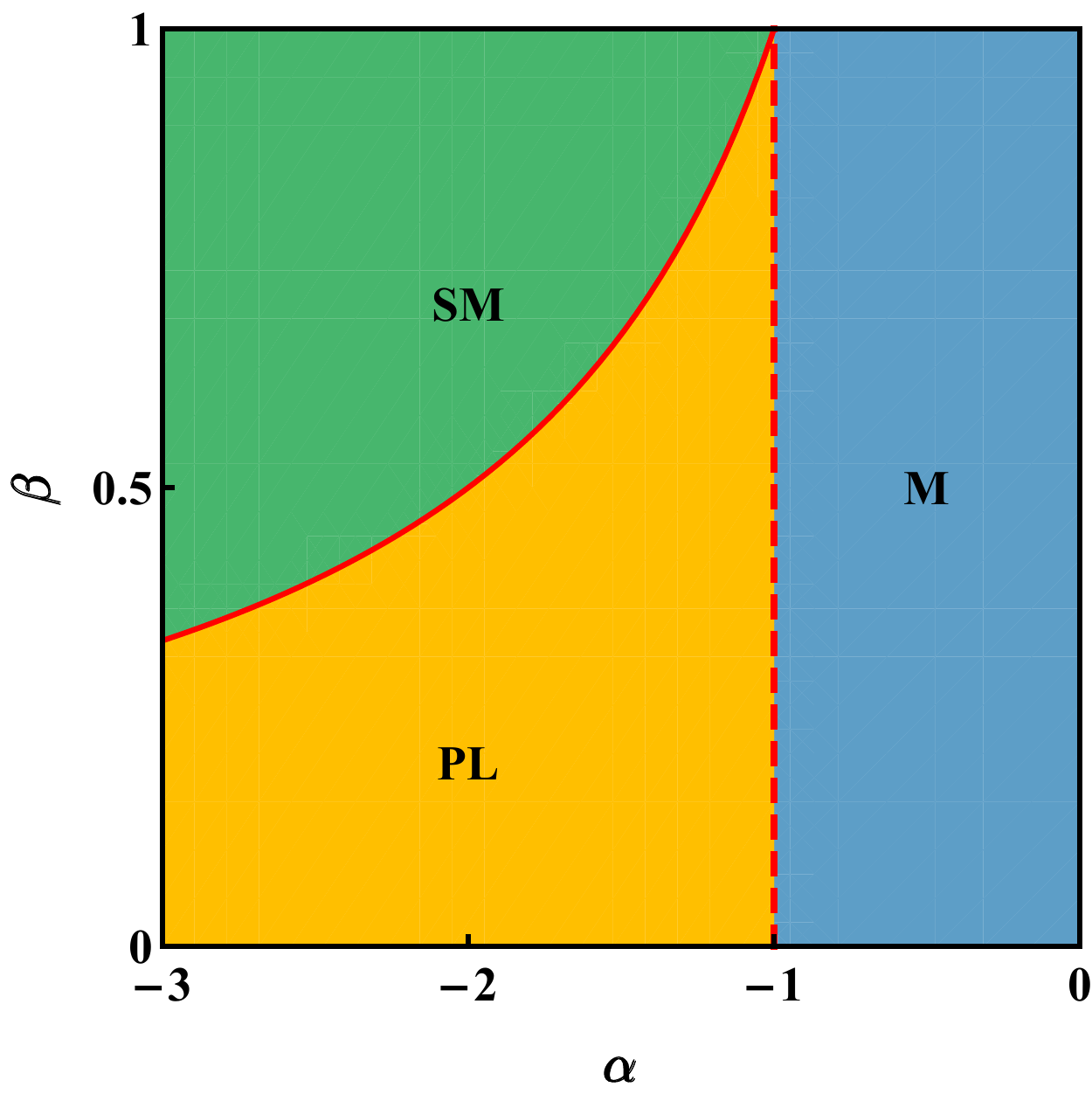}
\caption{Phase diagram of a bosonic binary mixture featuring \textit{attractive} interspecies coupling and confined in a generic $L-$site potential. Each of the three phases is associated to a specific functional dependence of the minimum-energy configuration $(\vec{x}_*,\vec{y}_*)$ and of $V_*$ (see relations (\ref{eq:V_*})) on parameters (\ref{eq:alfa_beta}). Phase M is the uniform and mixed one, phase PL features a soliton just in the minority species, while phase SM exhibits the presence of a supermixed soliton. Red dashed (solid) line corresponds to a phase transition where the first (second) derivative of $V_*$ with respect to $\alpha$ is discontinuous.  }
\label{fig:Diagramma_di_fase}
\end{figure}

The search for the configuration $(\vec{x},\vec{y})$ minimizing effective potential (\ref{eq:V}), on its domain $\mathcal{R}$ has been developed according to the fully-analytic scheme sketched in the previous Sec. and further illustrated in \cite{NoiSREP}. Interestingly, our analysis has highlighted the presence of a common phase diagram for systems featuring $L=2$ (dimer), $L=3$ (trimer) and $L=4$ (tetramer). Such a phase diagram is illustrated in Figure \ref{fig:Diagramma_di_fase} and includes three phases:

    \textit{i) Phase M (Mixed)} occurs for $\alpha >-1$ and features uniform boson distribution among the $L$ wells and mixing of the two species;

    \textit{ii) Phase PL (Partially Localized),} present for $\alpha<-1$ and $ \beta<-1/\alpha$, is such that the minority species, i.e. species b (since $N_b \sqrt{U_b}<N_a \sqrt{U_a}$), conglomerates and forms a soliton, while the majority species, i.e. species a, occupies all available wells, even if not in a uniform way;

    \textit{iii) Phase SM (SuperMixed)} is marked by the presence of a supermixed soliton (and full localization), meaning that both species conglomerate in the same well.

These three systems therefore feature a common pathway which, upon variation of control parameters $\alpha$ and $\beta$, leads from the uniform and mixed configuration (phase M) to the supermixed soliton (phase SM), through the intermediate phase (phase PL), characterized by partial localization, i.e already showing the seed of the soliton, whose emergence, in turn, is due to the localizing effect of the interspecies attraction. For this reason, we conjecture that the mechanism of formation of supermixed solitons is the same regardless of the value of $L$. To better connote the three presented phases, in Table \ref{tab:Phases} we give the explicit expressions of $(\vec{x}_*,\vec{y}_*)$ as functions of model parameters $\alpha$ and $\beta$, together with the relevant value of $V_*$ (recall relations (\ref{eq:V_*})), in each of the three phases.

\begin{figure}[h!]
\centering
\includegraphics[width=1\linewidth]{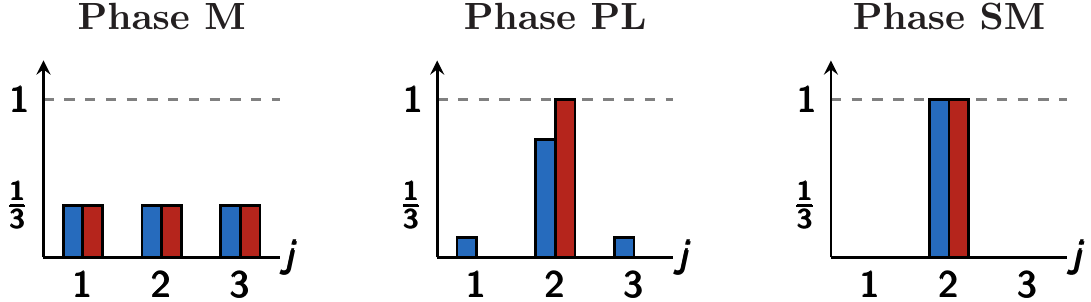}
\caption{Pictorial representation of the minimum-energy configurations for phases M, PL and SM, in a 3-well system. Vertical axis represent normalized populations $x_{*,j}$ and $y_{*,j}$ for the ground state, while numbers $1,\,2,\,3$ label the three wells. The majority (minority) species is depicted in blue (red) and corresponds to the left (right) columns of the histograms in each panel. In phase M the two species are uniformly distributed in the three wells; in phase PL the minority species forms a soliton while the majority species still occupies all available sites; in phase SM the interspecies attraction is so strong that a supermixed soliton is formed.}
\label{fig:Istogrammi}
\end{figure}

\begin{table}[]
\begingroup
\renewcommand{\arraystretch}{1.5} 
\begin{tabular}{|c|c|c|}
\hline
Phase & \(\displaystyle (\vec{x}_*,\vec{y}_*) \) & \(\displaystyle V_* \) \\ \hline
M & \begin{tabular}[c]{@{}c@{}}\(\displaystyle x_{*,j}=1/L \quad \forall j \)\\ \\ \(\displaystyle y_{*,j}=1/L \quad \forall j \)\end{tabular} & \(\displaystyle V_{*}^\mathrm{M}=\frac{1}{2L}(\beta^2+2\alpha\beta+1)\) \\ \hline
PL & \begin{tabular}[c]{@{}c@{}}\(\displaystyle x_{*,i}=[1-(L-1)\alpha\beta]/L \)\\  \(\displaystyle x_{*,j}=[1+\alpha\beta]/L  \quad \forall j\neq i \)\\ \\ \(\displaystyle y_{*,i}=1, \,\,\,  y_{*,j}=0 \,\,\, \forall j\neq i \)\end{tabular} & \begin{tabular}[c]{@{}c@{}}\(\displaystyle V_{*}^\mathrm{PL}=\frac{1}{2L} [1+2\alpha\beta\) \\ \(\displaystyle + \beta^2 (L-(L-1)\alpha^2)]\)\end{tabular} \\ \hline
SM & \begin{tabular}[c]{@{}c@{}}\(\displaystyle x_{*,i}=1\)\\  \(\displaystyle x_{*,j}=0 \quad \forall j\neq i \)\\ \\ \(\displaystyle y_{*,i}=1, \,\,\, y_{*,j}=0 \,\,\, \forall j\neq i\)\end{tabular} & \(\displaystyle V_{*}^\mathrm{SM}=\frac{1}{2}(\beta^2+2\alpha\beta+1)\) \\ \hline
\end{tabular}
\endgroup
\caption{Summary of the different functional dependencies of the minimum-energy configuration and of the relevant value of the effective potential (see relations (\ref{eq:V_*})) in each of the three phases.}
\label{tab:Phases}
\end{table}
We remark that the results listed in Table \ref{tab:Phases} have been derived in an analytic way (and numerically checked by means of a brute-force minimization of potential (\ref{eq:V})) for $L=2,\,3,\,4$ while it is quite natural to conjecture the validity of these results also for $L\ge 5$. To corroborate our conjecture, it is worth observing that, for any $L$, $V_*=V_*(\alpha,\beta)$ is continuous everywhere in the half-plane $\{(\alpha,\,\beta)\, :\, \alpha\le 0\,$ and $0\le\beta\le 1\}$. In particular, equations 
$$
V_{*}^\mathrm{M}(\alpha=-1,\beta)=V_{*}^\mathrm{PL}(\alpha=-1,\beta)
$$ 
and 
$$
V_{*}^\mathrm{PL}(\alpha,\beta=-1/\alpha)=V_{*}^\mathrm{SM}(\alpha,\beta=-1/\alpha)
$$
hold, respectively, at phase M-PL and phase PL-SM borders. On the other hand, one can easily realize that the \textit{first} derivative $\partial V_*/\partial\alpha$ is discontinuous at $\alpha=-1$ while the \textit{second} derivative $\partial^2 V_*/\partial\alpha^2$ is discontinuous at $\beta=-1/\alpha$, regardless of the specific value of $L$ (see the first panel of Figure \ref{fig:Grafici_classici}). This difference in the non-analyticity properties of $V_*$ at the two phase boundaries is a direct consequence of the specific functional dependence of $x_{*,j}$'s and $y_{*,j}$'s on model parameters $\alpha$ and $\beta$ in each of the three phases (see second column of Table \ref{tab:Phases}). The minimum energy configuration $(\vec{x}_*,\vec{y}_*)$, in fact, features a jump discontinuity at transition M-PL while it is continuous at transition PL-SM. In this regard, one can notice that $(\vec{x}_*,\vec{y}_*)$ exhibits the same $Z_L$ symmetry of the trapping potential just in phase M. By making the control parameter $\alpha$ more negative, one crosses the M-PL border and such symmetry \textit{suddenly} breaks. A soliton starts to emerge in a certain well, although the remaining $L-1$ wells still include part of the majority species (i.e. species a). Further increasing $|\alpha|$, the soliton emerges in a clearer and sharper way, since all the remaining wells are gradually emptied by the localizing effect of the interspecies attraction. At border PL-SM, the latter has become so strong that both species are fully localized in a certain well, leaving all the remaining ones empty: the supermixed soliton is now completely formed and a further increase of $|\alpha|$ has no effect on the minimum energy configuration $(\vec{x}_*,\vec{y}_*)$. This scenario is pictorially illustrated in Figure \ref{fig:Istogrammi} for the case $L=3$.    
We recall that generalized potentials (\ref{eq:Potenziale_lungo}) and (\ref{eq:V}) have been derived under the assumption that overall boson populations $N_a$ and $N_b$ are large enough (see Sec. \ref{subsec:CVP}). If this is not the case, the introduction of continuous variables is no longer legitimate and, for small or zero values of $T_a$ and $T_b$, the formation of the supermixed solitons will not occur in a continuous way with respect to the variation of a control parameter. On the contrary, in phase PL, the soliton will form and enlarge by incorporating one boson at a time. This phenomenology, whose inherently discretized essence is closely connected with the emergence of the Mott-insulator phase, will be discussed in a separate work.

\subsection{Entropy of mixing and Entropy of location as critical indicators} Two indicators that are well-known in Statistical Thermodynamics and Physical Chemistry \cite{Brandani,Camesasca}, the Entropy of mixing and the Entropy of location, can be conveniently used to detect the occurrence of phase transitions in the class of systems that we are investigating \cite{Noi_Zenesini}. They are, respectively defined as follows:
\begin{equation}
    \label{eq:S_mix}
    S_{mix}=-\frac{1}{2} \sum_{j=1}^L\left( x_j \log \frac{x_j}{x_j+y_j} +y_j \log \frac{y_j}{x_j+y_j} \right) 
\end{equation}
\begin{equation}
    \label{eq:S_loc}
    S_{loc}= -\sum_{j=1}^L \frac{x_j+y_j}{2} \log \frac{x_j+y_j}{2}.
\end{equation}
They provide complementary information about the degree of non-homogeneity present in the system. Namely, the former quantifies the degree of mixing while the latter measures the spatial localization of the particles irrespective of their species. 

\begin{figure}[h!]
\centering
\includegraphics[width=0.8\linewidth]{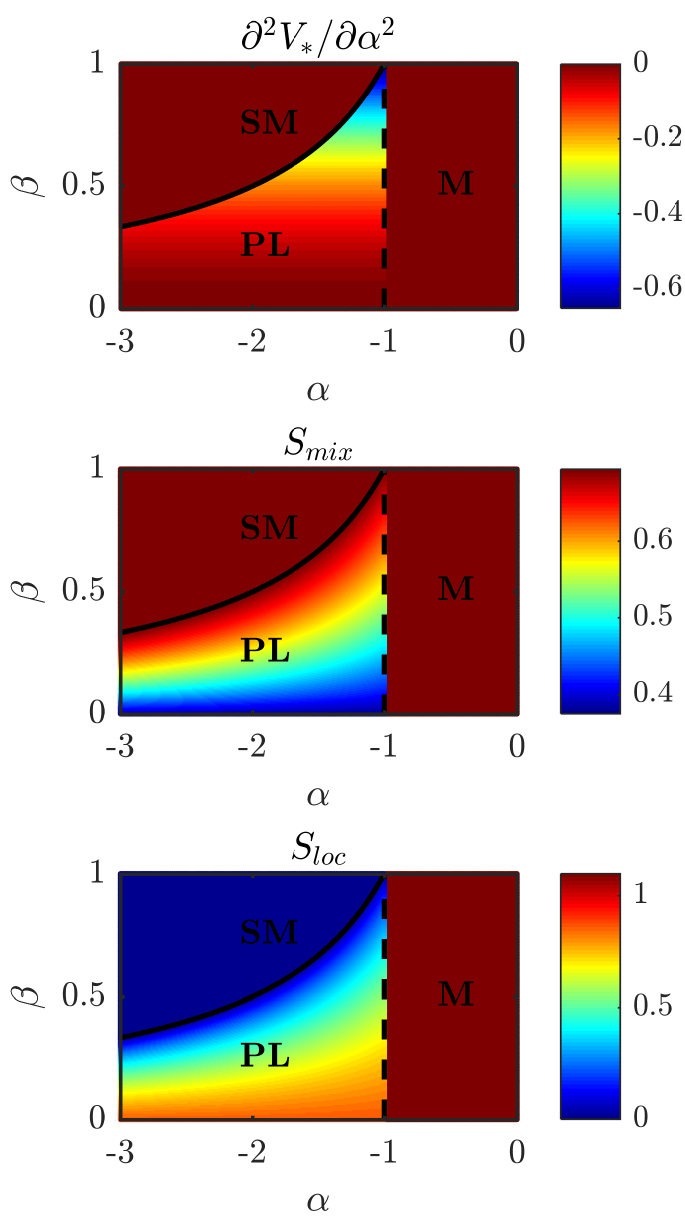}
\caption{Some critical indicators witnessing the presence of three different phases in an $(L=3)$-site potential (trimer) for $T_a/(U_aN_a)\to 0$ and $T_b/(U_bN_b)\to 0$. First panel: second derivative of functions $V_*^\mathrm{M}$, $V_*^\mathrm{PL}$ and $V_*^\mathrm{SM}$ (see third column of Table \ref{tab:Phases}) with respect to control parameter $\alpha$ for $L=3$. One can appreciate that it is discontinuous both at border PL-SM and at border M-PL (in the latter border the \textit{first} derivative $\partial V_*/\partial\alpha$ is already discontinuous). Second and third panel: critical indicators (\ref{eq:S_mix}) and (\ref{eq:S_loc}) associated to the minimum-energy configuration $(\vec{x}_*,\,\vec{y}_*)$ (obtained, in turn, setting $L=3$ in the second column of Table \ref{tab:Phases}). Note for the grayscale version: phase M, SM and the lower part of PL (first panel), M and SM (second panel) and M (third panel) correspond to the biggest value of the associated scale. The upper part of phase PL (first panel), the lower part of phase PL (second panel) and phase SM (third panel) correspond to the smallest value of the associated scale. }
\label{fig:Grafici_classici}
\end{figure}

By plugging the expressions of $x_{*,j}$'s and $y_{*,j}$'s associated to each of the three phases (see second column of Table \ref{tab:Phases}) into formulas (\ref{eq:S_mix}) and (\ref{eq:S_loc}), one can obtain particularly simple expressions for $S_{mix}$ and $S_{loc}$ in phase M and in phase SM, which read
$$
  S_{mix,\mathrm{M}}= \log 2, \qquad S_{loc,\mathrm{M}}=\log L,
$$
$$
  S_{mix,\mathrm{SM}}= \log 2, \qquad S_{loc,\mathrm{SM}}=0.
$$
Interestingly, $S_{mix}$ is the same both in phase M and in phase SM. This indicator, in fact, gives information just about the degree of mixing of the two atomic species, which is indeed the same both in the mixed and in the supermixed phase. Nevertheless, the profound difference between such phases can be appreciated by the combined use of $S_{mix}$ and $S_{loc}$, as the latter quantifies the degree of spatial delocalization of the atomic species among the wells. In phase PL, the analytic expressions of these indicators are rather complex (although straightforward to find) and, for the sake of clarity, we prefer to give their extreme values:
$$
   \min_{(\alpha,\beta)\in \text{PL}}S_{mix} =\frac{1}{2L}\left[L\log\left(1+\frac{1}{L}\right) + \log\left(1+L\right)\right] 
$$
$$
   \max_{(\alpha,\beta)\in \text{PL}}S_{mix} =\log 2 \equiv S_{mix,\mathrm{SM}}
$$
$$
\min_{(\alpha,\beta)\in \text{PL}}S_{loc} =  0 \equiv S_{loc,\mathrm{SM}}
$$
$$
\max_{(\alpha,\beta)\in \text{PL}}S_{loc} =\log(2L)-\frac{L+1}{2L}\log(L+1),
$$
which are found on the PL-SM border and on the line $\beta=0$.
The complete scenario on the $(\alpha,\beta)$-plane is illustrated (for $L=3$ sites) in the second and in the third panel of Figure \ref{fig:Grafici_classici}, where the presence of three qualitatively different regions is evident. 

\section{The delocalizing effect of tunnelling}
\label{sec:Delocalizing_effect}
\begin{figure}[h!]
\centering
\includegraphics[width=1\linewidth]{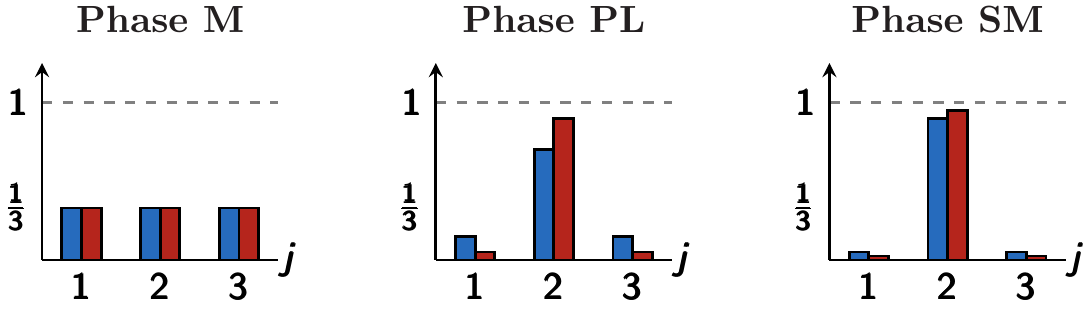}
\caption{Pictorial representation of the minimum-energy configurations for a 3-well system where the tunnelling processes are present. Vertical axis represent normalized populations $x_{*,j}$ and $y_{*,j}$ for the ground state, while numbers $1,\,2,\,3$ label the three wells. The majority (minority) species is depicted in blue (red) and corresponds to the left (right) columns of the histograms in each panel. Non-zero tunnelling processes determine the presence of residual tails at the two sides of the soliton but they do not significantly modify the scenario depicted in Figure \ref{fig:Istogrammi}.}
\label{fig:Istogrammi_T}
\end{figure}

As already mentioned, the presence of well-recognizable phases in the plane $(\alpha,\beta)$ sharply emerges when $T_a/(U_a N_a)\to 0$ and $T_b/(U_b N_b)\to 0$, two conditions that can be regarded as a sort of thermodynamic limit, according to the statistical-mechanical scheme developed in \cite{Dynamical_Bifurcation,Oelkers}. Moving away from these limits (either because the numbers of particles $N_a$ and $N_b$ are not large enough or because the hopping amplitudes $T_a$ and $T_b$ have a non-negligible weight in the overall energy balance of the system), the phase diagram illustrated in Figure \ref{fig:Diagramma_di_fase} and discussed in Sec. \ref{sec:Phase_diagram} gets smoothed and deformed, but it is still recognizable. The changes are essentially due to the delocalizing effect of tunnelling terms, which hinder the formation of localized configurations, i.e. of solitons (compare Figures \ref{fig:Istogrammi} and \ref{fig:Istogrammi_T}).

In a mathematical perspective, the presence of non-zero tunnelling terms has a regularizing effect on the generalized potential (\ref{eq:Potenziale_lungo}), whose global minimum can be determined with less effort than in the vanishing-tunnelling case, since such minimum always falls in the \textit{interior} of domain $\mathcal{R}$ and never on its \textit{boundary}. One therefore needs to look for the minimum-energy solution of equations $\nabla\mathcal{V}=0$, the gradient being computed with respect to the $2L-2$ independent variables $x_j$, $y_j$ where $j=1,\,2,\,\dots L-1 $ due to particle-number-conservation constraints.

\begin{figure}[h!]
\centering
\includegraphics[width=1\linewidth]{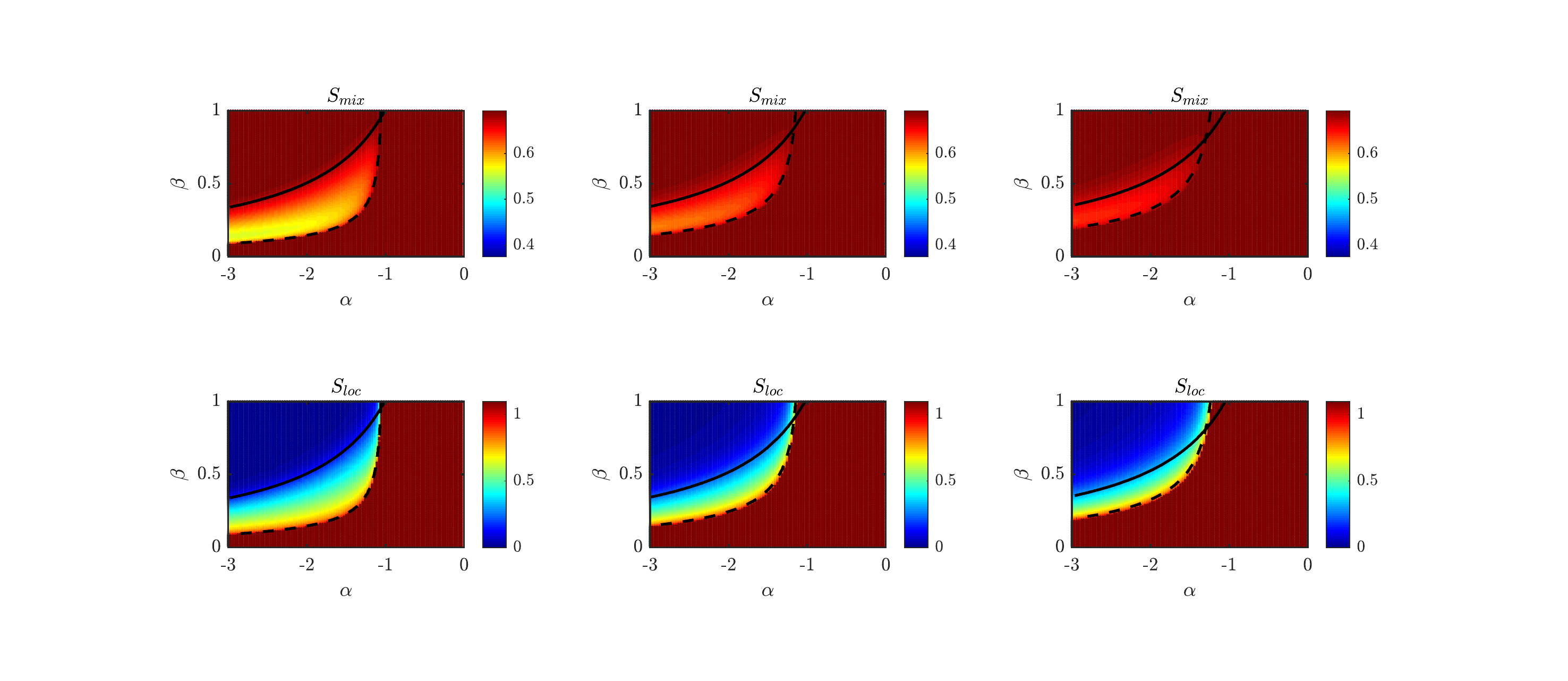}
\caption{Entropy of mixing and Entropy of location associated to the configuration $(\vec{x},\,\vec{y})$ minimizing potential (\ref{eq:Potenziale_lungo}), where $L=3$, $N_a=N_b=15$, $U_a=1$, $U_b\in[0,1]$ and $W\in[-3,0]$. From left to right, $T_a=T_b$ have been set, respectively, to $0.2$, $0.5$ and $0.8$. The dashed lines represent the border of the region where the uniform solution $x_j=y_j=1/3$ constitutes the minimum-energy configuration and where Bogoliubov frequencies, computed assuming the macroscopic occupation of a \textit{momentum} mode, are well defined. Their analytic expression is given by inequality (\ref{eq:Inequality_trimero}). The solid lines constitute the border of the region where Bogoliubov frequencies, computed assuming the macroscopic occupation of a \textit{site} mode, are well defined. Their analytical expression is given by formula (\ref{eq:Cond_Bogoliubov_solitone}). The comparison with Figure \ref{fig:Grafici_classici} shows that the phase diagram is modified by the presence of tunnelling processes, but it is not disrupted by them. Note for the grayscale version: in the first row, the darkest shade corresponds to the highest possible value of the associated scale; in (all panels of) the second row, moving from the upper left corner to the bottom right corner, the indicator continuously varies from the minimum to the maximum of the associated scale. }
\label{fig:Grafici_classici_T_non_0}
\end{figure}

We have fully developed this analysis for $L=2$ (dimer), $L=3$ (trimer) and $L=4$ (tetramer). Although we refer to Figure \ref{fig:Grafici_classici_T_non_0} (obtained setting $L=3$) for the sake of clarity, the following observations have been proved to hold for $L=2,\,3,\,4$ and are conjectured to be still valid also for $L\ge 5$:
\begin{itemize}
    \item Contrary to the zero-tunneling case, critical indicators $S_{mix}$ and $S_{loc}$ are continuous functions of model parameters $\alpha$ and $\beta$. This circumstance is due to the fact that normalized boson populations $x_j$'s and $y_j$'s themselves no longer feature jump discontinuities. Nevertheless, both indicators are still able to witness the presence of three qualitatively different regions in the $(\alpha,\,\beta)$ plane.
    \item Supported by tunneling processes, the mixed phase survives beyond the border $\alpha=-1$, provided that $\beta=N_b \sqrt{U_b}/( N_a\sqrt{U_a})$ is small enough. In this case, in fact, the interspecies attraction is hindered by the delocalizing effect of $T_a$ and $T_b$ so much that it is unable to trigger soliton formation. Interestingly, by resorting to the Hessian matrix associated to effective potential (\ref{eq:Potenziale_lungo}), it is possible to derive  inequality
    \begin{equation}
        \label{eq:Inequality_trimero}
        \alpha >-\sqrt{\left(1+\frac{9}{2}\frac{T_a}{U_aN_a}\right)\left(1+\frac{9}{2}\frac{T_b}{U_bN_b}\right)}
    \end{equation}
    giving the region of parameters' space where the uniform configuration is the least energetic one, i.e. where the configuration $x_j=y_j=1/3$ represents not only \textit{a local} but also \textit{the global} (constrained) minimum of function (\ref{eq:Potenziale_lungo}).  This region, whose border is depicted with dashed lines in Figure \ref{fig:Grafici_classici_T_non_0}, coincides (in the limit $N_a=N_b$, $T_a=T_b$, $U_a=U_b$) with the portion of parameters' space where Bogoliubov quasi-particle frequencies are well defined \cite{NoiPRA2} (we remark that such spectrum was computed assuming the macroscopic occupation of a \textit{momentum} mode).
    
    \item The formation of a supermixed soliton, the configuration for which $S_{loc}= 0$, is only slightly hindered by the presence of tunnelling processes. The latter tend to delocalize the atomic species among the wells and are responsible for the survival of non-zero tails in wells far from the supermixed soliton. Nevertheless, such tails, which are fully reabsorbed by the soliton only in the limit $\alpha \to -\infty$, do not significantly affect the solitonic structure of the minimum-energy configuration (see third panel of Figure \ref{fig:Istogrammi_T}). This circumstance is witnessed by the fact that, in the upper left part of the phase diagram, $S_{loc}$ is only slightly lower than $\log L $ (see second row of Figure \ref{fig:Grafici_classici_T_non_0}).  

\end{itemize}

\onecolumngrid

\twocolumngrid

With reference to Figure \ref{fig:Grafici_classici_T_non_0}, we remark that, along the dashed lines (representing the border between phase M and phase PL and given by formula (\ref{eq:Inequality_trimero})), the Bogoliubov frequencies computed assuming the macroscopic occupation of a \textit{momentum} mode vanish \cite{NoiPRA2}. Conversely, along the solid lines (representing the border between phase PL and phase SM and given by formula (\ref{eq:Cond_Bogoliubov_solitone})), the Bogoliubov frequencies computed assuming the macroscopic occupation of a \textit{site} mode vanish (see Appendix \ref{sec:App_Bogoliubov_solitone}).  

\subsection{Uniform configuration for a generic $L$-site potential}
It is possible to analytically derive the counterpart of inequality (\ref{eq:Inequality_trimero}), which holds for $L=3$, both for the dimer ($L=2$) and for the tetramer ($L=4$). These inequalities, ensuing from the condition that the Hessian matrix associated to generalized potential (\ref{eq:Potenziale_lungo}) and evaluated at point $x_j=y_j=1/L$ is positive definite, respectively read
\begin{equation}
        \label{eq:Inequality_dimero}
        \alpha >-\sqrt{\left(1+2\frac{T_a}{U_aN_a}\right)\left(1+2\frac{T_b}{U_bN_b}\right)}
\end{equation}
and
\begin{equation}
        \label{eq:Inequality_tetramero}
        \alpha >-\sqrt{\left(1+4\frac{T_a}{U_aN_a}\right)\left(1+4\frac{T_b}{U_bN_b}\right)}.
\end{equation}
It is worth mentioning that their twin-species limits (i.e. their expression when $N_a\to N_b$, $U_a \to U_b$ and $T_a \to T_b$) coincide with the inequalities giving the regions of parameters' space where Bogoliubov quasi-particle frequencies are well defined. The latter have been derived, assuming the macroscopic occupation of a \textit{momentum} mode, for the dimer in \cite{PennaLinguaPRE} and in \cite{NoiPRA2}, thanks to the dynamical algebra method, for a ring lattice. In view of these results and of the rather general formulas giving the condition for the collapse of Bogoliubov frequencies in a generic $(L\ge 3)$-site ring lattice (see \cite{NoiPRA2}), it is quite natural to conjecture that, for a generic $L$-site potential and for $T_a\neq T_b$, $U_a\neq U_b$ and $N_a\neq N_b$, inequality
\begin{equation}
        \label{eq:Inequality_L}
        \alpha >-\sqrt{\left[1+C_L\frac{T_aL}{U_aN_a}\right]\left[1+C_L\frac{T_bL}{U_bN_b}\right]},
\end{equation}
where $C_L=1-\cos(2\pi/L)$, gives the region of parameters' space where the uniform solution $x_j=y_j=1/L$ is the least energetic one. Conversely, going out of region (\ref{eq:Inequality_L}), the uniform solution ceases to be a local (and also the global) minimum of function (\ref{eq:Potenziale_lungo}), a circumstance which corresponds to the onset of the transition between phase M and phase PL. Remarkably, in the limit $T_a/(U_a N_a) \to 0$ and $T_b/(U_b N_b)\to 0$, inequalities (\ref{eq:Inequality_trimero}), (\ref{eq:Inequality_dimero}), (\ref{eq:Inequality_tetramero}) and (\ref{eq:Inequality_L}) reduce to $\alpha>-1$, the condition which was shown to constitute the border between phase M and PL in the thermodynamic limit (see Figure \ref{fig:Diagramma_di_fase}). In passing, one can observe that, for $L=2$, the mismatch between inequalities (\ref{eq:Inequality_L}) and (\ref{eq:Inequality_dimero}) is only apparent, in that the former is referred to a system inherently featuring the ring geometry which is absent in the dimer.

\section{Quantum critical indicators}
\label{sec:Quantum_indicators}
\begin{figure}[h]
\centering
\includegraphics[width=1\linewidth]{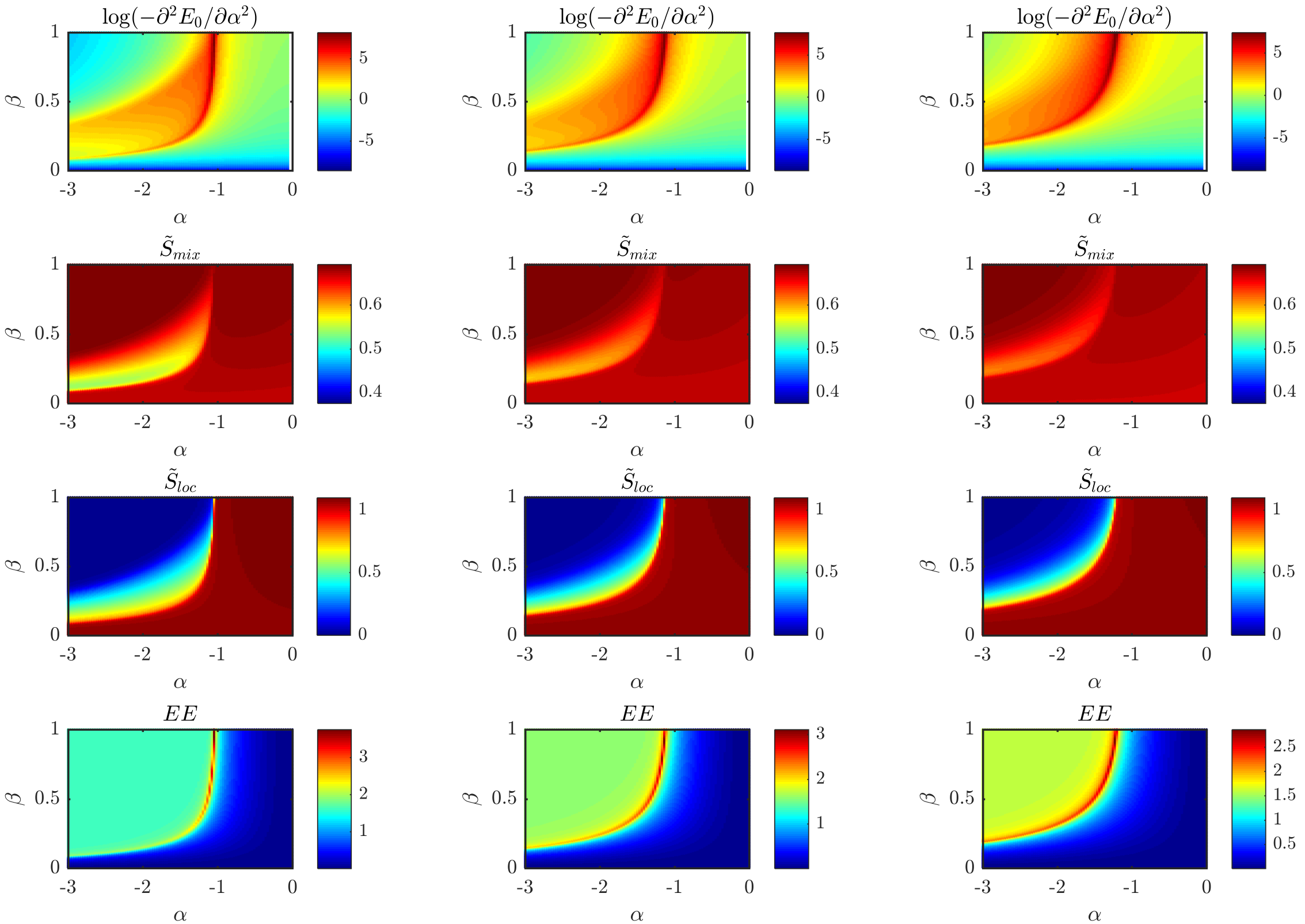}
\caption{Each row illustrates the behaviour of a genuinely quantum indicator as a function of model parameters $\alpha$ and $\beta$. Each column corresponds to a different value of the ratio $T/U_a$, where $T:=T_a=T_b$ (from left to right, $T/U_a=0.2,\,0.5,\,0.8$). First row: second derivative of the ground-state energy $E_0$ (see formula \ref{eq:Ground_state_energy}) with respect to $\alpha$. The logarithmic scale is used in order to better visualize the presence of three qualitatively different regions. Second row: quantum version of the entropy of mixing, $\tilde{S}_{mix}$ (see formula \ref{eq:S_mix_tilde}). Third row: quantum version of the entropy of location $\tilde{S}_{loc}$ (see formula \ref{eq:S_loc_tilde}). Fourth row: entanglement between the two condensed species, $EE$ (see formula (\ref{eq:EE})). Model parameters $L=3$, $N_a=N_b=15$, $U_a=1$, $U_b\in[0,1]$ and $W\in[-3,0]$ have been used. Note for the grayscale version: in the first row, the darkest shade in the central and the bottom regions correspond to the maximum and to the minimum, respectively, of the associated scale; second row: the darkest shade corresponds to the highest possible value of the associated scale; in (all panels of) the third row, moving from the upper left corner to the bottom right corner, the indicator continuously varies from the minimum to the maximum of the associated scale; fourth row: the indicator assumes the minimum (maximum) possible value in the right and in the lower part (along an hyperbolic-like line in the vicinity of the M/PL transition) of each panel.}
\label{fig:Quantum_indicators_trimero}
\end{figure}

The mechanism of formation of supermixed solitons presented in Sec. \ref{sec:Phase_diagram} and \ref{sec:Delocalizing_effect} by means of a semiclassical approach capable of highlighting, in a rather transparent way, the presence of three different phases in the plane $(\alpha,\beta)$, is fully confirmed by genuinely quantum indicators. To develop the quantum analysis, one has to perform the exact numerical diagonalization \cite{HPC_Polito} of Hamiltonian (\ref{eq:BH}) in order to determine the ground state 
\begin{equation}
\label{eq:Ground_state}
   |\psi_0\rangle = \sum_{\vec{n},\,\vec{m}}^Q c(\vec{n},\,\vec{m}) |\vec{n},\vec{m}\rangle,
\end{equation}
the associated energy
\begin{equation}
\label{eq:Ground_state_energy}
   E_0=\langle \psi_0 |H|\psi_0\rangle
   \end{equation}
and the first excited levels
\begin{equation}
\label{eq:Livelli_eccitati}
   E_i=\langle \psi_i |H|\psi_i\rangle.
\end{equation}
Of particular importance for the current investigation are coefficients $c(\vec{n},\vec{m})$ appearing in expansion (\ref{eq:Ground_state}) and defined as 
\begin{equation}
\label{eq:c_k}
    c(\vec{n},\vec{m})= \langle\vec{n},\vec{m}|\psi_0\rangle 
\end{equation}
which will be used to introduce the quantum counterparts of indicators (\ref{eq:S_mix}) and (\ref{eq:S_loc}). The diagonalization of Hamiltonian (\ref{eq:BH}) is carried out for extended sets of model parameters, in such a way to explore vast regions of the $(\alpha,\,\beta)$-plane (recall formulas (\ref{eq:alfa_beta})), also in relation with the presence of non-negligible hoppings $T_a$ and $T_b$. This analysis allows one to appreciate the dependence of some genuinely quantum indicators on model parameters and, above all, their being critical along the same curves of the $(\alpha,\,\beta)$-plane where the semiclassical approach predicts the occurrence of mixing-supermixing transitions. For the sake of clarity, we will refer to Figure \ref{fig:Quantum_indicators_trimero}, whose rows correspond to different quantum indicators and whose columns to different values of the hopping amplitude $T:=T_a=T_b$. Going from left to right, it reads 
\begin{equation}
    \label{eq:T}
    T=0.2,\, 0.5,\, 0.8
\end{equation}
respectively. In general, the same observations that we made in Sec. \ref{sec:Delocalizing_effect} concerning the delocalizing effect of tunnelling and the impact thereof on $S_{mix}$ and on $S_{loc}$, hold also within this purely quantum scenario. In particular, one can notice that: \textit{i)} All quantum indicators are continuous functions of model parameters $\alpha$ and $\beta$, \textit{ii)} The mixed phase is supported by tunnelling processes, \textit{iii)} The formation of supermixed solitons occurs for large values of $|\alpha|$ and moderate values of $\beta$. 

The quantum critical indicators which have been scrutinized in relation to the mixing-supermixing transitions are the following:

\textit{Ground-state energy.} Observing indicator (\ref{eq:Ground_state_energy}), regarded as a function of effective model parameters $\alpha$ and $\beta$, one can appreciate the presence of three different phases (corresponding to the already discussed phase M, phase PL and phase SM). To be more clear, function $E_0(\alpha,\,\beta)$ is everywhere continuous in the $(\alpha,\,\beta)$-plane, but it features non analiticities, either in its first or in its second derivative, along two specific lines of the phase diagram which, in turn, divide the latter into three separate regions. The functional dependence of $E_0$ in each of the three regions is different, that means that the slope $\partial E_0 /\partial \alpha$ and the concavity  $\partial^2 E_0 /\partial \alpha^2$ exhibit different behaviours.

This circumstance is well illustrated in the first row of Figure \ref{fig:Quantum_indicators_trimero}, where we have plotted $\partial^2 E_0 /\partial \alpha^2$ (the logarithmic scale has been adopted just for graphical purposes) for three different values of the hopping amplitude. The left panel, obtained for $T/U_a=0.2$, allows one to recognize two regions (in green), well separated by an intermediate region (in red-orange) which intercalates between them. In the central and in the right panels, which feature bigger hopping amplitudes ($T/U_a=0.5$ and $0.8$ respectively), the presence of the intermediate phase (phase PL) is still evident, although it turns out to be slightly deformed and its borders less sharp. 

\textit{Entropy of mixing.} In Sec. \ref{sec:Phase_diagram} we introduced indicator (\ref{eq:S_mix}) and discussed its ability to quantify the degree of mixing of a semiclassical configuration $(\vec{x},\,\vec{y})$. A reasonable quantum mechanical version of this indicator can be constructed as follows: after determining the complete decomposition (\ref{eq:Ground_state}) of the system's ground state $|\psi_0\rangle$ and, in particular, the full list of coefficients (\ref{eq:c_k}) (the cardinality of this set being given by formula (\ref{eq:Q})), one can evaluate the entropy of mixing of $|\psi_0\rangle$ by defining
\begin{equation}
\label{eq:S_mix_tilde}
   \tilde{S}_{mix}:=\sum_{\vec{n},\vec{m}}^Q |c(\vec{n},\vec{m})|^2 S_{mix}(\vec{n},\vec{m}),
\end{equation}
where $S_{mix}(\vec{n},\vec{m})$ is the entropy of mixing of the state $(\vec{n},\vec{m})$ of the Fock basis, computed by means of formula (\ref{eq:S_mix}) (with the obvious identifications $x_j=n_j/N_a$ and $y_j=m_j/N_b$). 

The indicator thus obtained is illustrated, as a function of model parameters $\alpha$ and $\beta$, in the second row of Figure \ref{fig:Quantum_indicators_trimero} for the three choices (\ref{eq:T}). Especially for small hoppings, one can observe the presence of an intermediate phase (phase PL) which stands in between phase SM and phase M. Increasing the tunnelling, the inter-phase borders tend to get less sharp and the distinction between the phases gets decreasingly evident. Interestingly, the results given by quantum indicator (\ref{eq:S_mix_tilde}), whose employment requires the knowledge of the full list of coefficients (\ref{eq:c_k}), are in very good agreement with those ones obtained within the CVP (compare the panels in the first row of Figure \ref{fig:Grafici_classici_T_non_0} with the corresponding ones in the the second row of Figure \ref{fig:Quantum_indicators_trimero}, obtained for the same model parameters). 

\textit{Entropy of location.} With a similar reasoning, one can define the quantum counterpart of classical indicator (\ref{eq:S_loc}), i.e.
\begin{equation}
\label{eq:S_loc_tilde}
   \tilde{S}_{loc}:=\sum_{\vec{n},\vec{m}}^Q |c(\vec{n},\vec{m})|^2 S_{loc}(\vec{n},\vec{m}),
\end{equation}
where coefficients $c(\vec{n},\vec{m})$ are given by formula (\ref{eq:c_k}) and $S_{loc}(\vec{n},\vec{m})$ is the entropy of location associated to the state $(\vec{n},\vec{m})$ of the Fock basis and computed by means of formula (\ref{eq:S_loc}) (with the obvious identifications $x_j=n_j/N_a$ and $y_j=m_j/N_b$). The behaviour of indicator $\tilde{S}_{loc}$ in the $(\alpha,\,\beta)$-plane is illustrated in the third row of Figure \ref{fig:Quantum_indicators_trimero}. In the three panels corresponding to values (\ref{eq:T}), similar to the  case of $\tilde{S}_{mix}$, it is possible to identify phase M (in red), phase SM (in blue), and the intermediate one (where $\tilde{S}_{loc}$ varies between $\approx 0$ and $\approx \log L =\log 3$).

Its remarkable specificity and sensitivity, together with the non-small extent of its range, make this indicator particularly suitable for the detection of soliton-like configurations. It is worth mentioning that the results obtained within a purely quantum treatment (i.e. numerically diagonalize Hamiltonian (\ref{eq:BH}), obtain coefficients (\ref{eq:c_k}) and plug them into formula (\ref{eq:S_loc_tilde})) well match those obtained within the semiclassical CVP approach (compare the panels in the second row of Figure \ref{fig:Grafici_classici_T_non_0} with the corresponding ones in the third row of Figure \ref{fig:Quantum_indicators_trimero}, which share the same model parameters).

\textit{Entropy of Entanglement (EE).} The degree of quantum correlation between two partitions of its can effectively mirror the structure of a given ground state $|\psi_0\rangle$,  which, in turn, can radically change upon variation of model parameters \cite{PennaLinguaJPB,NoiEntropy,NoiSREP}. Among various possibilities, we have focused on the entropy of entanglement between species a and species b. As a consequence, the entanglement between the two atomic species is given by 
\begin{equation}
        EE = - \mathrm{Tr}_{a} (\hat{\rho}_{a}\, \log_2 \hat{\rho}_{a} ), 
\label{eq:EE}
\end{equation}
an expression corresponding to the Von Neumann entropy of the reduced density matrix 
\begin{equation}
     \hat{\rho}_{a}=\mathrm{Tr}_{b} \left(\hat{\rho}_0\right).
\label{eq:Reduced_density_matrix}
\end{equation}
The latter can be obtained, in turn, by tracing out the degrees of freedom of species b from the ground state's density matrix $\hat{\rho}_0=|\psi_0\rangle\langle\psi_0|$. The fourth row of Figure \ref{fig:Quantum_indicators_trimero} illustrates indicator $EE$ as a function of $\alpha$ and $\beta$ for the three values (\ref{eq:T}). 

One can notice that, when $\alpha\to 0$, then $EE\to 0$ since in this limit the two species do not interact. Increasing $|\alpha|$, $EE$ features a sharp peak exactly where the transition between phase M and phase PL takes place, a circumstance which has been already noticed in relation to mixing-demixing transitions \cite{PennaLinguaJPB,NoiEntropy,NoiSREP}. Further increasing $|\alpha|$, a plateau is reached, wherein the $EE$ stabilizes to the limiting value of $\log L=\log_2 3 \approx 1.59$. The argument of the logarithm (which is set to $L=3$ in the example shown in Figure \ref{fig:Quantum_indicators_trimero}), corresponds to the number of semiclassical configurations minimizing potential (\ref{eq:V}) and which are quantum-mechanically reabsorbed in the formation of a unique non-degenerate ground state. In other words, the $L$-fold degeneracy of the semiclassical configuration corresponding to the presence of a supermixed soliton in one of the $L$ wells is lifted by the presence of tunnelling, which therefore determines the formation of a $L$-faced Schr\"{o}dinger cat.

\begin{figure}[h]
\centering
\includegraphics[width=1\linewidth]{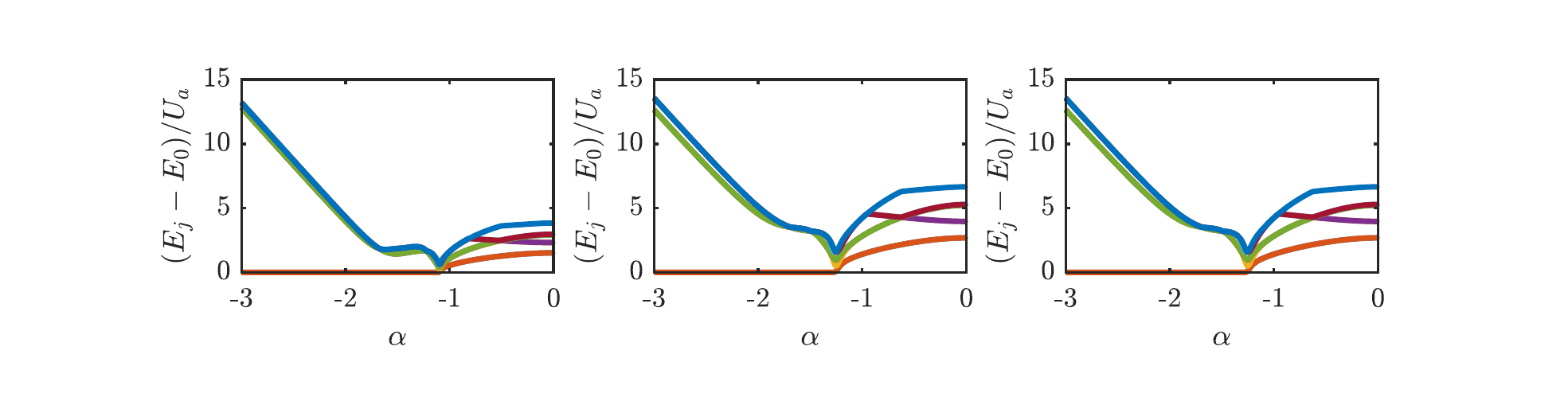}
\caption{ First 8 excited energy levels, obtained by means of an exact numerical diagonalization of Hamiltonian (\ref{eq:BH}), for a $L=3$-system and for $T:=T_a=T_b=0.2,\,0.5,\,0.8$ in the left, central and right panel, respectively. Model parameters $N_a=N_b=15$, $U_a=1$, $U_b=0.36 $ and $W\in[-1.8,0]$ have been chosen. }
\label{fig:Spectrum_trimero}
\end{figure}

\textit{Energy spectrum.} The computation of the first excited energy levels of the system (see formula (\ref{eq:Livelli_eccitati})) as a function of control parameter $\alpha$ can give an additional physical insight and a further confirmation of the presence of three qualitatively different phases. Figure \ref{fig:Spectrum_trimero} illustrates the energy fingerprint of a $L=3$ system, for $\beta = 0.6$ and the usual values (\ref{eq:T}). With reference to the left panel, the one featuring the smallest value of $T/U_a$, it is possible to distinguish three different regions wherein the energy-levels arrangement is qualitatively different. For small values of $|\alpha|$, the levels can be shown to well match Bogoliubov's quasi-particles frequencies which are, in turn, computed assuming the macroscopic occupation of momentum mode $k=0$ (see \cite{NoiPRA2}). At $\alpha \approx -1$ all these levels collapse, thus signing the end of phase M and, further increasing $|\alpha|$ they manifestly rearrange (it is worth mentioning that, for $\alpha<-1$ some excited levels seem to coincide with the lowest one, but, actually, this overlap is just apparent and merely due to the scale used for the vertical axis).  Further increasing $|\alpha|$ down to $\alpha \approx -1.7$, another qualitative change of the energy levels' structure is met, which constitutes the border between phase PL and phase SM. At such value of $\alpha$, in fact, the energy levels, although they do not collapse, assume a distinctly-linear functional dependence on $\alpha$. The presence of three regions where the energy fingerprint is qualitatively different can be noticed also in the central and in the right panel of Figure \ref{fig:Spectrum_trimero}, although the critical behaviours (namely the spectral collapse and the onset of the linear ramp) are smoothed down by the delocalizing effect of tunnelling. In this regard, one can observe that tunnelling is responsible also for the leftward translation of the collapse point (see formula (\ref{eq:Inequality_trimero}) and the discussion thereof). 

\onecolumngrid


\twocolumngrid

\section{Concluding remarks}
\label{sec:Conclusions}

In this work, we have investigated the mechanism of soliton formation in bosonic binary mixtures loaded in ring-lattice potentials. Our analysis has evidencded that all these systems, irrespective of the number sites, share  a common mixing-demixing phase diagram. The latter is spanned by two effective parameters, $\alpha$ and $\beta$, the first one representing the ratio between the interspecies attraction and the (geometric average of) the intraspecies repulsions, the second one accounting for the degree of asymmetry between the species. Such phase diagram includes three different regions, differing in the degree of mixing and localization. The first phase, occurring for sufficiently small $|\alpha|$, is the mixed one (phase M) and it is such that the atomic species are perfectly mixed and uniformly distributed among the wells. The second phase (phase PL) occurs for moderate values of $|\alpha|$ and sufficiently asymmetric species. It includes the seed of localized soliton-like states, although the latter are not developed in a full way. Eventually, the third phase (phase SM), occurring for sufficiently large values of $|\alpha|$, corresponds to states such that both atomic species clot in the same unique well, hence the name supermixed solitons.

After introducing the quantum model and its representation in the CVP, in Sec. \ref{sec:Phase_diagram}, the mixing-supermixing transitions are derived within such semiclassical approximation scheme which transparently shows  the  emergence of a bi-dimensional phase diagram. The three phases therein not only feature specific functional dependences of the ground-state energy on model parameters, but also  are characterized in terms of two critical indicators imported from Statistical Thermodynamics, the entropy of mixing and the entropy of location.

Sec. \ref{sec:Delocalizing_effect} is devoted to the analysis when the ratio $T/(UN)$ is small but non-zero, i.e. how the phase diagram changes and gets blurred if one walks away from the thermodynamic limit (in the sense specified within the statistical mechanical approach developed in \cite{Dynamical_Bifurcation,Oelkers}). The delocalizing effect of tunneling is shown to favor the mixed phase and to hinder the formation of solitons but not to upset the presented phase diagram. Quantum indicators are presented in Sec. \ref{sec:Quantum_indicators}, whose critical behaviour along certain lines of the phase diagram $(\alpha,\,\beta)$ corroborates the scenario that emerged from the semiclassical treatment of the problem.

In conclusion, we note that the methodology on which our analysis relies, together with the classical and quantum indicators used to detect critical phenomena, can be easily applied to systems with more complex lattice topologies, interactions and tunnelling processes \cite{Jason_stella,Viscondi, Chianca,Penna_Cavaletto,Salasnich_dipolari}. In view of this, and considering the increasing interest for multicomponent condensates \cite{Ytterbium,Eto_Nitta,Spin_turbulence,Hartman_drag}, our future work will aim to extend the presented analysis to the soliton formation's mechanism in complex lattices and in presence of multiple condensed species.

\begin{appendices}
\section{}
\label{sec:App_Bogoliubov_solitone}
In this appendix, we derive, by means of a modified version of the Bogoliubov approximation scheme \cite{NoiPRA1,NoiPRA2}, the analytical expression of quasiparticles' frequencies of a $L=3$-system when its ground state exhibits a supermixed soliton-like structure (namely, when it belongs to phase SM). In this circumstance, in fact, one can recognize that there are two site modes, $a_1$, $b_1$, that are macroscopically occupied, namely $ n_1\approx N_a -n_2-n_3$ and $m_1\approx N_b -m_2-m_3$ while the microscopically occupied ones are $a_2$, $a_3$, $b_2$ and $b_3$. With these substitutions in mind, one can derive $H^{(2)}$, the quadratic approximation of the original Hamiltonian (\ref{eq:BH}), which reads
$$
   H^{(2)} \approx -T_a(a_3^\dagger a_2+ a_2^\dagger a_3  ) -(U_a N_a+N_bW)(n_2+n_3)  
$$
$$
    -T_b ( b_3^\dagger b_2+ b_2^\dagger b_3  ) -(U_b N_b+N_aW)(m_2+m_3)  .
$$
Notice that we have neglected not only higher-order terms but also linear terms, since the latter contribute just to the ground-state energy but do not affect the characteristic frequencies and, in general, they can be removed by a suitable unitary transformation.

\begin{figure}[ht]
    \centering
    \includegraphics[width=0.7\linewidth]{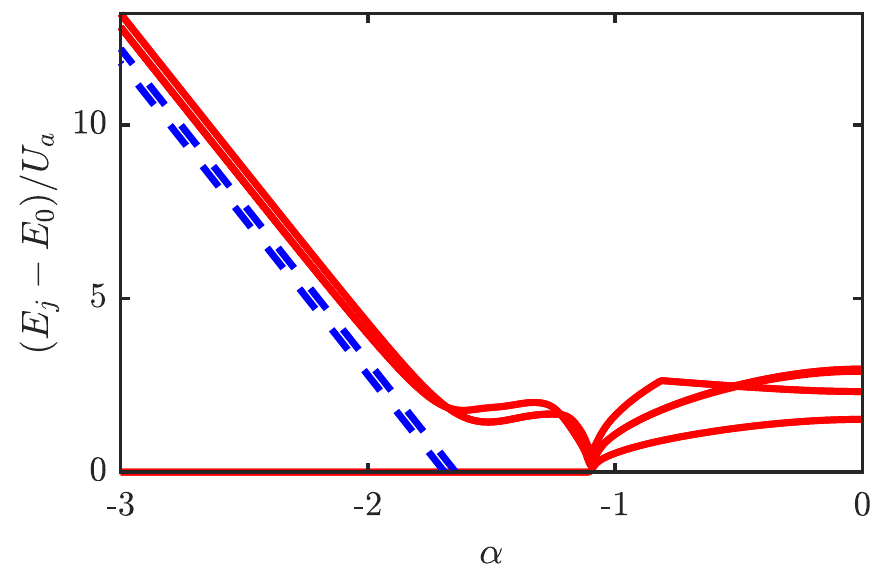}
    \caption{Red solid lines: first excited levels of the exact spectrum obtained by means of numerical diagonalization of Hamiltonian (\ref{eq:BH}). Blue dashed lines: Bogoliubov characteristic frequencies present in diagonal Hamiltonian $H_D$. The following model parameters have been chosen: $L=3$, $T_a=T_b=0.2$, $U_a=1$, $U_b=0.36$, $N_a=N_b=15$, $W\in[-1.8,\,0]$.  }
    \label{fig:Confronto_Bogoliubov_numerico}
\end{figure}

Recognizing that terms 
$$
   J_+ = a_2a_3^\dagger, \qquad J_-= a_2^\dagger a_3, \qquad J_3=\frac{1}{2}(n_3-n_2)
$$
constitute the two-boson realization of algebra su(2), one can easily diagonalize $H^{(2)}$ enacting the unitary transformation $U_\varphi= e^{\frac{\varphi}{2}(J_+-J_-)} $  which gives
$$
  U_\varphi (J_++J_-)  U_\varphi^\dagger =   2 J_3 \sin\varphi + (J_++J_-) \cos\varphi.
$$
Treating in the same way terms $b_j$'s, it is straightforward to derive diagonal Hamiltonian 
$$
    H_D= n_2(T_a-U_a N_a-N_b W) + n_3(-T_a-U_a N_a-N_b W)
$$
$$
    +m_2(T_b-U_b N_b-N_a W) +m_3(-T_b-U_b N_b-N_a W),
$$
an expression where the coefficients of number operators constitute the Bogoliubov quasiparticles' frequencies, namely $H_D= \omega_2 n_2 +\omega_3 n_3 + \Omega_2 m_2 +\Omega_3 m_3$. As illustrated in Figure \ref{fig:Confronto_Bogoliubov_numerico}, the agreement between the spectrum envisaged by this approximation scheme and the exact one, obtained numerically, is good, not only qualitatively (same linear behaviour) but also quantitatively ($<10\%$ of difference if $|\alpha|$ is large enough). This agreement rapidly improves as soon as the numbers of particles $N_a$ and $N_b$ increase.

Interestingly, the simultaneous validity of conditions
\begin{equation}
    \label{eq:Cond_Bogoliubov_solitone}
    \omega_2>0,\quad \omega_3 >0,  \quad  \Omega_2>0, \quad  \Omega_3 >0
\end{equation}
gives the region of paramaters' space where Hamiltonian $H_D$ is lower bounded, i.e. the region where the supermixed soliton-like configuration is estimated to be stable. The border of this region corresponds to the solid lines present in Figure \ref{fig:Grafici_classici_T_non_0} which, in turn, stand where indicators $S_{mix}$ and $S_{loc}$ illustrated therein feature criticalities.

In conclusion, we remark that the approximation scheme developed in this appendix is based on the assumption of macroscopic occupation of \textit{site} modes (one for each component) and that it is able to estimate the energy spectrum for large values of $|\alpha|$, i.e. in phase SM. This scheme is therefore fundamentally different from the one developed in \cite{NoiPRA2} and linked to condition (\ref{eq:Inequality_trimero}), since the latter was based on the assumption of macroscopic occupation of momentum mode $k=0$ and was therefore intended to approximate the energy spectrum for small values of $|\alpha|$ (a circumstance corresponding, in turn, to uniform boson configuration, i.e. to phase M). 

\end{appendices}

\end{document}